\documentclass[a4paper]{cas-dc}

\usepackage[utf8]{inputenc}
\usepackage[numbers]{natbib}

\usepackage{graphicx}
\usepackage{caption}
\usepackage{amssymb}
\usepackage{amsmath}
\usepackage{xcolor}

\usepackage{lineno}
\usepackage{array}
\usepackage{microtype}

\usepackage{cleveref} 
\hypersetup{
    colorlinks=true,  % 启用颜色链接
    linkcolor=green,  % 设置链接颜色为绿色
    citecolor=green   % 设置引用颜色为绿色（可选）
}

\crefname{figure}{Figure}{Figs.}
\Crefname{figure}{Figure}{Figs.}
\begin{document}
\let\WriteBookmarks\relax
\def\floatpagepagefraction{1}
\def\textpagefraction{.001}
\shorttitle{DualPhys-GS}
\shortauthors{Li et~al.}

\title [mode = title]{DualPhys-GS: Dual Physically-Guided 3D Gaussian Splatting for Underwater Scene Reconstruction}                      
% \tnotemark[1]

\tnotetext[1]{This work is supported by project ZR2024QF286 supported by Shandong Provincial Natural Science Foundation, National Key Research and Development Program of China (NO. 2022YFB4004401), National Natural Science Foundation of China (NO. 62202425), the Taishan Scholars Program (NO. tsqnz20240834), the Qilu Youth Innovation Team (NO. 2024KJH028)}

\author[1,2]{Jiachen Li}[orcid=0000-0002-3543-6088]

% \cormark[1]
% \fnmark[1]
\ead{jcli@qlu.edu.cn}
% \ead[url]{www.jkkrishnan.in}

% \credit{Conceptualization of this study, Methodology, Software}

%\address[1]{, Street 129, 1043 NX Amsterdam, The Netherlands}
\affiliation[1]{organization={Key Laboratory of Computing Power Network and Information Security, Ministry of Education},
                addressline={Shandong Computer Science Center (National Supercomputer Center in Jinan)},
                city={Jinan},
%               citysep={}, % Uncomment if no comma needed between city and postcode
                % postcode={}, % 如果有邮编可以在这里填写，没有就保持为空
                % state={Shandong},
                country={China}}

\author[1,2]{Guangzhi Han}
\ead{10431240234@stu.qlu.edu.cn}
% \credit{Data curation, Writing - Original draft preparation}

\author[1,2]{Jin Wan}
% [%
%    role=Co-ordinator,
%    suffix=Jr,
%    ]
% \fnmark[2]
\ead{wanj@qlu.edu.cn}
% \ead[URL]{https://www.university.org}

% \credit{Data curation, Writing - Original draft preparation}

\affiliation[2]{organization={Shandong Provincial Key Laboratory of Computing Power Internet and Service Computing, Shandong Fundamental Research Center for Computer Science},
                % addressline={}, % 如果没有额外的地址行，保持为空
                city={Jinan},
%               citysep={}, % Uncomment if no comma needed between city and postcode
                % postcode={}, % 如果有邮编可以在这里填写，没有就保持为空
                % state={Shandong},
                country={China}}

\author[1,2]{Yuan Gao}
% \cormark[2]
% \fnmark[1,3]
\ead{yuangao@qlu.edu.cn}
% \ead[URL]{www.campus.in}
% \credit{Data curation, Writing - Original draft preparation}

\author[1,2]{Delong Han}
\cormark[1]
% \fnmark[1,3]
\ead{handl@qlu.edu.cn}
% \ead[URL]{www.campus.in}

\cortext[cor1]{Corresponding author}
% \cortext[cor2]{Principal corresponding author}
% \fntext[fn1]{This is the first author footnote, but is common to third
%   author as well.}
% \fntext[fn2]{Another author footnote, this is a very long footnote and
%   it should be a really long footnote. But this footnote is not yet
%   sufficiently long enough to make two lines of footnote text.}

% \nonumnote{This note has no numbers. In this work we demonstrate $a_b$
%   the formation Y\_1 of a new type of polariton on the interface
%   between a cuprous oxide slab and a polystyrene micro-sphere placed
%   on the slab.
%   }

\begin{abstract}
In 3D reconstruction of underwater scenes, traditional methods based on atmospheric optical models cannot effectively deal with the selective attenuation of light wavelengths and the effect of suspended particle scattering, which are unique to the water medium, and lead to color distortion, geometric artifacts, and collapsing phenomena at long distances. We propose the DualPhys-GS framework to achieve high-quality underwater reconstruction through a dual-path optimization mechanism. Our approach further develops a dual feature-guided attenuation-scattering modeling mechanism, the RGB-guided attenuation optimization model combines RGB features and depth information and can handle edge and structural details.\ In contrast, the multi-scale depth-aware scattering model captures scattering effects at different scales using a feature pyramid network and an attention mechanism. Meanwhile, we design several special loss functions. The attenuation scattering consistency loss ensures physical consistency.\ The water body type adaptive loss dynamically adjusts the weighting coefficients.\ The edge-aware scattering loss is used to maintain the sharpness of structural edges.\ The multi-scale feature loss helps to capture global and local structural information. In addition, we design a scene adaptive mechanism that can automatically identify the water-body-type characteristics (e.g., clear coral reef waters or turbid coastal waters) and dynamically adjust the scattering and attenuation parameters and optimization strategies. Experimental results show that our method outperforms existing methods in several metrics, especially in suspended matter-dense regions and long-distance scenes, and the reconstruction quality is significantly improved.
\end{abstract}

% \begin{graphicalabstract}
% \includegraphics[width=\textwidth]{pipeline.png}
% \end{graphicalabstract}

% \begin{highlights}
% \item Decouple underwater optical effects into RGB-guided attenuation and multi-scale depth-aware scattering models
% \item Enforce physical consistency through specialized loss functions that maintain attenuation-scattering relationships
% \item Automatically adapt to diverse water environments through adaptive scene classification and dynamic parameter adjustment
% \end{highlights}


\maketitle

\section{Introduction}
The importance of 3D reconstruction techniques for underwater scenes is increasing with the growing demand for marine resource exploitation and ecological monitoring.\ However, compared with the atmospheric environment, the available data for underwater scenes are limited by acquisition costs and equipment conditions, and face unique physical challenges: (1) Wavelength selective attenuation, the water body absorbs light of different wavelengths with significant differences, and in particular, long-wavelength light (e.g., red light) attenuates drastically with distance, resulting in color distortion (typically blue-green hue), which affects the color fidelity of the rendered image. (2) Multiple scattering effect, Anisotropic scattering effect triggered by suspended particles in water (e.g., plankton, organic debris and minerals).\ In long-distance or turbid water bodies, forward scattering makes the geometric structure blurred, background noise more significant, and background scattering makes the image `foggy', which reduces the accuracy of feature matching.\ (3) Diversity of water environment, different water bodies exhibit different optical properties, making it difficult for fixed parameter models to maintain robustness in different water environments.\ Traditional reconstruction methods based on atmospheric optical models (e.g.,\ MVS \cite{furukawa2015} ) in underwater environments not only suffer from systematic deviations in geometric accuracy, but also suffer from significant degradation in texture fidelity, seriously restricting the application potential of underwater 3D reconstruction technology in the fields of marine science, autonomous navigation of underwater robots, and virtual underwater tours.

Neural Radiance Fields (NeRFs) \cite{milden2020} have demonstrated their unique advantages in underwater scenes in recent years, which effectively address some of the challenges in underwater scenes by modeling implicit neural radiation fields in conjunction with body rendering techniques. By accumulating densities and colors along the light rays, volume rendering can accurately simulate the translucency effect in the water body and represent the complex optical interactions between underwater objects and suspended particles at different depths.\ In addition, the volume-integrated nature of volume rendering helps to simulate the propagation process of light rays in different aqueous media, further supporting the modeling of wavelength-dependent light transport. However, NeRF-based methods \cite{barron2021mipnerf} \cite{bian2023nope} \cite{tang2024neural} \cite{ramazzina2023scatternerf} are usually accompanied by high computational overheads during training and rendering, and cannot realize real-time rendering.\ Meanwhile, during global optimization, it is difficult to accurately distinguish the effects of wavelength-selective attenuation in body rendering, leading to color distortion at long distances, and these limitations constrain its potential application in real-time underwater scene reconstruction.

Recently, 3D Gaussian Splatting (3DGS) \cite{kerbl2023} based on explicit Gaussian representations has made a breakthrough in 3D scene reconstruction.\ 3DGS \cite{kerbl2023} innovatively introduces explicit 3D Gaussian representations and a differentiable rasterization pipeline, improving real-time rendering efficiency while maintaining visual quality.\ In addition, its anisotropic Gaussian primitives accurately model the geometric details of the scene, effectively preserving the edge structure of the underwater scene, and its transparency-based alpha blending mechanism facilitates the integration of physically-based color recovery models. However, the atmospheric medium-based light transport assumption of 3DGS \cite{kerbl2023} faces a fundamental limitation when directly migrated to underwater scenes: it ignores the unique optical effects of the water column. The wavelength-dependent attenuation of light by the water column leads to a systematic color bias of distant objects (especially a substantial loss of red information). Anisotropic scattering effects induced by suspended particles interfere with the depth-consistent estimation, leading to biased reconstruction of the density field and geometric structure. These problems are manifested in the 3D reconstruction results as geometric artifacts in the distant seafloor topography, systematic color bias in the coral reef texture, and pseudo-volumetric effect on the scene surface.

To solve these problems, a 3DGS-based dual-path optimization framework is proposed.\ DualPhys-GS combines with a dual-physics modeling scene adaptation method to achieve high-quality reconstruction of underwater scenes. The core innovation of DualPhys-GS lies in the design of a feature-guided attenuation-scattering dual-modeling mechanism based on the feature-guided attenuation-scattering mechanism, which accurately decomposes underwater optical propagation processes into two key physical processes, namely, attenuation and scattering. For the attenuation process, we design an RGB-guided attenuation optimization model, which combines RGB features and depth information to process the scene edges and structural details accurately, and introduces the wavelength physics a priori, which simulates the water body's differentiation of the absorption of light at different wavelengths.\ For the scattering process, we present a multi-scale depth-aware scattering model, which captures the scattering effects at different scales through a feature pyramid network and an attention mechanism. To ensure that the reconstruction results conform to the underwater optical laws, we design special loss functions such as edge-aware scattering loss, multi-scale feature loss, and attenuation-scattering consistency loss,\ guaranteeing the system's high-quality reconstruction capability in various underwater environments. In addition, we realize a scene adaptive mechanism for the diversity of underwater environments, which can automatically identify the type of water body and dynamically adjust the optimization parameters and strategies. In Summary, Our key contributions are as follows:
\begin{enumerate}
\item We propose a feature-guided attenuation-scattering dual modeling mechanism based on an RGB-guided attenuation optimization model and  multi-scale depth-aware scattering model to achieve accurate simulation of underwater optical propagation processes.
\item We design edge-aware scattering loss, multi-scale feature loss, attenuation scattering consistency loss, and water body type adaptive loss to ensure that the model output conforms to the laws of underwater optical physics.
\item We propose a scene-adaptive mechanism that automatically recognizes the type of water body and dynamically adjusts the optimization strategy to ensure high-quality reconstruction in different underwater scenes.
\end{enumerate}
\section{Related work}
%% Inline mathematics is tagged between $ symbols.
\subsection{Underwater reconstruction based on conventional multi-view geometry}
Traditional underwater 3D reconstruction mainly relies on Multi-View Stereo (MVS) \cite{furukawa2015}.\ To address the optical interference of the water medium, researchers propose physical model-based compensation methods.\ Chambah et al.\ \cite{chambah2004} apply the Jaffe-McGlamery light transport equation to underwater scenes, estimate the background scattered light through a background scattering model, and combine it with color correction to recover the target reflectivity.\ However, this method is highly dependent on idealized water body assumptions and performs erratically in real environments with significant parameter variations.\ On the other hand, Drews-Jr et al.\ \cite{drews2013} estimate the transmittance field by using the Dark Channel Prior (DCP) \cite{he2010single}, which effectively suppresses the ambiguity effect caused by forward scattering. Nevertheless, these methods rely on accurate calibration of the optical parameters of the water body (e.g., attenuation coefficient, scattering phase function) and have poor robustness in turbid waters. In recent years, deep learning methods enter this field. Li et al.\ \cite{li2018} design a two-branch network to estimate the medium parameters and scene depth separately, which alleviates the dependence on physical priori parameters, but the scarcity of supervised data and the diversity of water-body-types make it difficult to generalize the model to underwater environments with different optical properties.

Based on the above studies, traditional multi-view stereo vision (MVS) \cite{furukawa2015} underwater reconstruction methods still face four fundamental challenges. Firstly, the unique optical properties of water bodies invalidate the assumptions of traditional algorithms, which leads to significant degradation of underwater image feature matching quality. This degradation becomes particularly severe in long-distance or turbid regions where feature mismatch rates remain high.\ Secondly, physical model compensation methods generally suffer from parameter sensitivity issues.\ These methods require tedious manual parameter adjustments for different water environments and lack self-adaptive mechanisms.\ Thirdly, existing approaches struggle to simultaneously handle the dual effects of wavelength-selective attenuation and multiple scattering effects.\ Their performance particularly deteriorates in complex scenarios with drastic water type variations. Finally, learning-based methods face limitations due to the high cost of underwater data acquisition and labeling difficulty.\ The scarcity of training data severely restricts their generalization capability. These collective issues cause traditional MVS methods \cite{akkaynak2018revised} \cite{huo2021efficient} in practical underwater applications to exhibit three persistent artifacts: long-range geometric collapse, blurred edge structures, and systematic color distortion.
\subsection{Underwater reconstruction based on NeRFs}
Neural Radiation Fields (NeRFs)\ \cite{milden2020} achieve high-quality 3D reconstruction through implicit neural representation.\ Their powerful implicit representation capability and volume rendering mechanism provide new solutions for modeling complex optical phenomena in underwater scenes, where current approaches mainly follow two directions,\ physical model enhancement and optical effects decoupling. Nevertheless, NeRF-based underwater reconstruction still faces critical challenges caused by medium scattering.

In the physical model enhancement direction, WaterNeRF \cite{Sethuraman2023} pioneers the integration of the Beer-Lambert attenuation law with the volume rendering equation, successfully simulating wavelength-dependent light attenuation.\ This physically-grounded formulation introduces rigorous radiative transfer constraints, significantly improving color fidelity in long-range underwater scenes.\ However, its neglect of scattering effects limits its ability to address background radiation interference from suspended particles.

For optical effects decoupling, Seathru-NeRF \cite{levy2023} achieves breakthrough by explicitly modeling scattering medium properties through light path decomposition.\ It enhances reconstruction quality in turbid media by separating direct transmission from scattered radiance.\ Despite this advancement, the method relies on oversimplified assumptions of globally homogeneous scattering coefficients, failing to account for spatially varying water medium parameters.\ This limitation becomes particularly pronounced in complex water bodies with dynamic optical properties.

The optical effects decoupling strategy manifests in two representative approaches, Ye et al.'s underwater light field preservation method \cite{ye2022} and WaterHE-NeRF \cite{zhou2023}. The former implicitly models forward scattering through joint optimization of the radiation field and optical transmission paths. However, its reliance on isotropic phase function approximations creates inherent limitations in characterizing anisotropic scattering properties of suspended particles in turbid waters.\ The latter innovatively decouples water refraction, surface fluctuation, and medium scattering into separate implicit fields. While this decomposition enhances physical interpretability, the computational complexity escalates exponentially, severely constraining rendering efficiency and making real-time applications currently infeasible.

Recent advancements in Beyond NeRF Underwater \cite{beyondnerf2023} push the boundaries further by implementing differentiable ray tracing for joint optimization of target reflectivity and medium parameters. This co-optimization framework achieves state-of-the-art rendering accuracy but introduces critical dependencies on dense depth sensor data. Consequently, the method struggles in monocular or sparse-view scenarios where depth information remains incomplete or unreliable. The above methods commonly suffer from coupled optimization of the background radiation field and the target reflectivity, leading to color bias in the long-range reconstruction. 
\subsection{Underwater reconstruction based on 3DGS}
3D Gaussian Splatting (3DGS) \cite{kerbl2023} provides a new paradigm for real-time underwater reconstruction with the advantages of explicit Gaussian characterization and differential rasterization. Numerous enhancement methods have been proposed to address challenging problems across various domains. These approaches encompass anti-aliasing \cite{song2024hdgs} \cite{condor2025dont}, deblurring \cite{wu2024deblur4dgs} \cite{oh2024deblurgs} \cite{dai2024spikenvs}, relighting \cite{jiang2024gaussianshader} \cite{li2024recap}, sparse view \cite{huang2025fatesgs} \cite{cai2024dust}, and diffusion models \cite{lin2025diffsplat} \cite{asim2025met3r}.

In underwater scene reconstruction, WaterSplatting \cite{li2024watersplatting} pioneers the adaptation of 3DGS \cite{kerbl2023} by incorporating learnable wavelength-selective attenuation coefficients, but it retains NeRF volumetric rendering.\ This approach effectively compensates for long-range color distortion while maintaining real-time efficiency, but its oversimplified physical model completely neglects scattering effects, leading to blurred edges and detail loss in turbid environments.\} To address this limitation, SeaSplat \cite{yang2024} introduces a comprehensive physical imaging model that embeds Henyey-Greenstein phase functions within Gaussian attributes for single-scattering approximation. Although this significantly improves reconstruction quality in turbid waters, three critical challenges persist: the single-scale representation fails to resolve near-field details and far-field structural features, the absence of multi-scattering and background radiation modeling restricts physical accuracy, and the globally fixed scattering residual parameters lack adaptability across diverse water types from clear reefs to turbid coasts.

Recent advancements explore advanced parameterization strategies. Aquatic-GS \cite{liu2024aquatic} implements hybrid implicit-explicit characterization to predict spatially varying medium parameters,\ enhancing adaptability to complex optical properties. However, its implicit component introduces additional computational overhead, compromising the inherent real-time advantage of 3DGS. Meanwhile, Gaussian Splashing \cite{mualem2024} focuses on dynamic effects through Gaussianized water volume modeling, successfully simulating surface waves but neglecting geometric accuracy optimization for static underwater structures.UW-GS\cite{wang2025uwgs} introduces a color appearance model and a physically-based density control mechanism that focuses on dynamic object processing and the generation of binary motion masks. The method handles underwater optical effects through color appearance modeling and employs physical density control to constrain the Gaussian distribution. Although innovative in dealing with dynamic scenes, it still has limitations in dealing with complex underwater optical phenomena. These developments collectively highlight the ongoing tension between physical accuracy, computational efficiency, and scenario adaptability in underwater 3DGS applications.

Current underwater scene reconstruction methods face three interlinked fundamental limitations rooted in physical modeling and computational framework design.\ The inherent coupling between attenuation and scattering parameters constitutes a primary challenge, as underwater light propagation simultaneously involves wavelength-dependent absorption and multi-scale scattering phenomena.\ In turbid environments dominated by suspended particles, existing frameworks systematically overestimate attenuation coefficients while underestimating scattering effects, whereas in clear waters this parameter imbalance reverses. Such coupled miscalibrations induce cumulative errors over long propagation paths, manifesting as chromatic distortions and geometric collapse of distant structures.

Furthermore, the optical heterogeneity across aquatic environments, from the crystalline waters of coral reefs to sediment-laden coastal zones, exposes critical adaptation limitations. Fixed parameter configurations optimized for specific water types fail to maintain performance consistency when deployed in non-target environments, necessitating manual recalibration during cross-scenario applications and severely constraining operational flexibility.\ Compounding these issues, the prevalent single-scale representation paradigm proves inadequate for capturing multi-range optical interactions.\ Near-field regions require high-frequency detail preservation to resolve particulate scattering effects, while far-field reconstruction demands robust structural coherence maintenance.\ Current single resolution frameworks either over smooth proximate details or discard critical background features, particularly failing to represent the gradual transition of scattering characteristics across depth-varying water columns.
\section{Preliminaries}
\subsection{3D Gaussian Splatting}
3D Gaussian Splatting (3DGS) \cite{kerbl2023} is an explicit 3D reconstruction scene representation that models a 3D scene by expanding the point cloud into anisotropic Gaussian primitives. Each Gaussian primitive $G_{i}$ consists of a centre position $\boldsymbol{\mu_{i}}$, a covariance matrix $\boldsymbol{\Sigma_{i}}$, and color attributes of opacity $\alpha_{i}$ and viewpoint. The density distribution function of Gaussian primitives is defined as follows:
%% Displayed equations can be tagged using various environments. 
%% Single line equations can be tagged using the equation environment.
\begin{equation}
G_{i}(x) = \mathrm{e}^{-\frac{\mathrm{1}}{\mathrm{2}}(x - \boldsymbol{\mu_{i}})^\mathrm{T} \boldsymbol{\Sigma_{i}}^{-\mathrm{1}} (x - \boldsymbol{\mu_{i}})}
\end{equation}
where $\boldsymbol{\mu_{i}} \in \mathbb{R}^{\mathrm{3}}$ denotes the spatial coordinates of the centre of the ellipsoid, and the covariance matrix $\boldsymbol{\Sigma_{i}}=\boldsymbol{R_{i}}\boldsymbol{S_{i}}^{\mathrm{2}}\boldsymbol{R}^{\mathrm{T}}_{i}$ denotes the shape and orientation of the Gaussian ellipsoid. The 3D Gaussian is projected to the 2D plane along any viewpoint. The projected 2D Gaussian maintains the Gaussian distribution characteristics. The Jacobian matrix $\boldsymbol{J}$ can compute its covariance matrix on the image plane, which can satisfy the demand of real-time rasterization. Then, alpha blending is performed to realize the semi-transparent effect through the viewpoint of depth-ordered Gaussian primitives. Given the pixel coordinates $x$ on the image plane, its final color $C_{i}(x)$ is calculated as follows:
\begin{equation}
C_{i}(x) = \sum_{i=\mathrm{1}}^{N} \boldsymbol{c}_i \alpha_i(x) \prod_{j=\mathrm{1}}^{i-\mathrm{1}} (\mathrm{1} - \alpha_j(x))
\end{equation}
where the color term $\boldsymbol{c}_{i}$ is encoded by the spherical harmonic function (SH), while the density property $\alpha_{i}$ is derived from Gaussian covariance, collectively defining the $i$-th Gaussian's photometric and geometric attributes.
\subsection{Underwater physical imaging model}
When reconstructing underwater scenes, the effects of water as a medium need to be considered compared to atmospheric media. Light propagating underwater undergoes the interaction of attenuation and backscattering effects, leading to light absorption during underwater propagation and, ultimately, to unnatural colors and artefacts in the rendered image. Classical underwater image formation \cite{akkaynak2018revised} can be expressed as a linear superposition of directly transmitted light and background scattered light with the following expression:
\begin{equation}
\boldsymbol{I}(x) = \boldsymbol{J}(x) \cdot T_{\mathrm{D}} + B_{\infty} \cdot (\mathrm{1} - T_{\mathrm{B}})
\end{equation}
where $\boldsymbol{I}$ is the irradiance received by the pixel $x$ out of the sensor, $\boldsymbol{J}$ denotes the color when there is no water attenuation, $T_{\mathrm{D}} = \mathrm{exp} \left( -\beta_d(x) \cdot z(x) \right)$ is the direct transmittance, $B_{\infty}$ denotes the background at infinite distance, $T_{\mathrm{B}} = \mathrm{exp} \left( -\beta_b(x) \cdot z(x) \right)$ denotes the background transmittance, and is correlated with the depth information $z$.

\section{Method}
\subsection{Overviwe of Dualphys-GS}

\begin{figure*}[!t]%% placement specifier
%% Use \includegraphics command to insert graphic files. Place graphics files in 
%% working directory.
\centering%% For centre alignment of image.
\includegraphics[width=\textwidth]{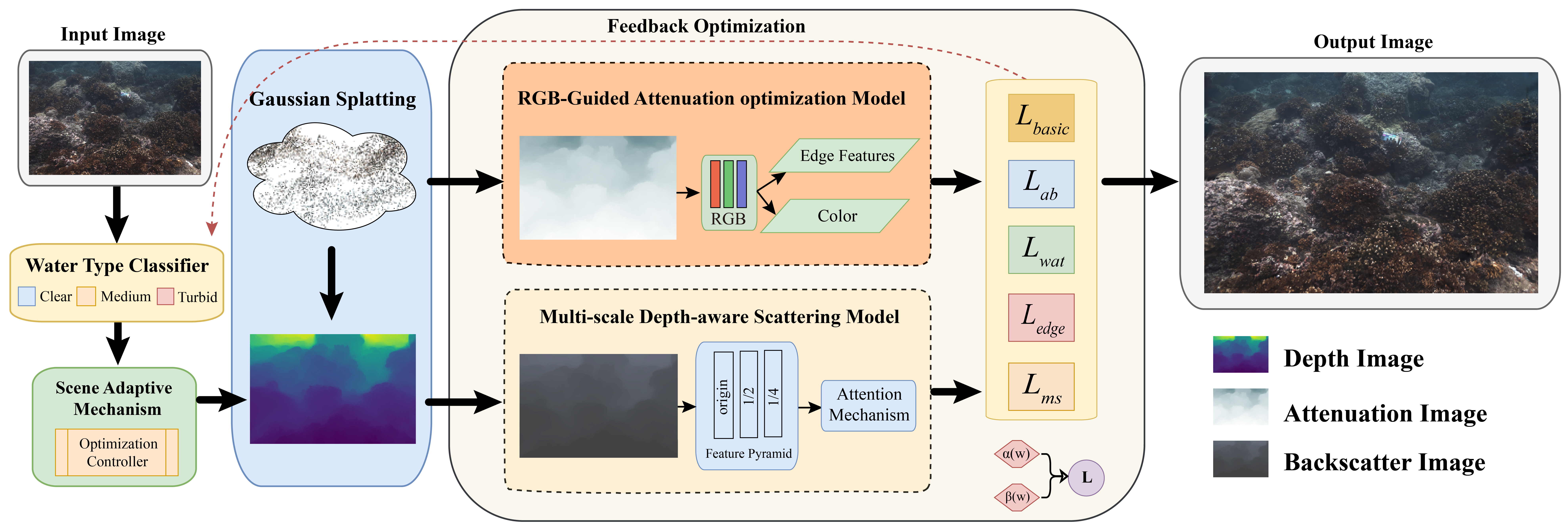}
%% Use \caption command for figure caption and label.
% \captionsetup{justification=raggedright, singlelinecheck=false}
\caption{\textbf{The Pipeline of DualPhys-GS.} We integrate the enhanced underwater physical imaging model with the 3DGS framework. Our water body adaptation mechanism first automatically identifies different scenarios through RGB inputs, followed by a processing pipeline comprising, 3DGS-based scene reconstruction, dual-path physical modeling with an RGB-guided attenuation model and multi-scale depth-aware scattering model, and ultimately mutual constraint optimization through loss functions, where $\alpha$ and $\beta$ denote parameter weights, $w$ represents water type categories, and $L$ constitutes the compound loss function.}\label{fig:fig1}
%% https://en.wikibooks.org/wiki/LaTeX/Importing_Graphics#Importing_external_graphics
\end{figure*}
As shown in \Cref{fig:fig1},\ we devise a physics-driven dual-path framework to constrain 3D Gaussian Splatting (3DGS) \cite{kerbl2023} for underwater scene reconstruction, which requires decoupling the original scene radiance $\boldsymbol{J}$ from water medium effects through three coordinated stages.\ The input image first undergoes water-type classification to automatically configure environmental adaptation parameters, followed by Structure-from-Motion (SfM) \cite{schonberger2016structure} processing to initialize point clouds and camera poses. In the attenuation path, an RGB feature extraction network collaborates with the attenuation model to characterize wavelength-dependent attenuation properties, while in the scattering path, a feature pyramid network enhances the scattering model's capacity to represent multi-scale anisotropic scattering patterns.\ These physically-derived attenuation and scattering estimates are synthesized through a radiative transfer model to generate final renderings. The training process incorporates four complementary loss functions, the consistency loss $L_{\mathrm{ab}}$ to ensure the physical correlation between attenuation and scattering, the water body adaptive loss $L_{\mathrm{wat}}$ to dynamically adjust the  weighting coefficients, the edge-aware loss $L_{\mathrm{edge}}$ to maintain structural clarity, and the multi-scale feature loss $L_{\mathrm{ms}}$ to capture structural information at different scales. 
\subsection{A dual attenuation-scattering modelling mechanism based on feature guidance}
The optical properties of underwater environments are far more complex than those of atmospheric environments and stem primarily from the effect of water as a medium on light propagation.\ As light passes through a body of water, it is affected by two main physical phenomena, light attenuation and backscattering. Light attenuation causes color distortion of objects at long distances, and backscattering introduces a “foggy” effect in the image, resulting in poor reconstruction. Traditional underwater physical imaging models are simple, decomposing the underwater image into a combination of attenuated direct light and scattered light.\ However, this simplification cannot effectively deal with underwater scenes' complexity, especially with edges, depth-varying regions, and different water conditions.\ To address this problem, we propose a feature-guided dual attenuation-scattering-based modeling mechanism, which accurately decomposes the underwater optical propagation process into two key physical processes, attenuation and scattering, and models them separately by feature enhancement. The underwater image formation model we adopted can be represented as:
\begin{equation}
\boldsymbol{I} = \boldsymbol{J} \cdot \boldsymbol{A}(\boldsymbol{D}, \boldsymbol{I}_{\mathrm{RGB}}) + \boldsymbol{B}(\boldsymbol{D})
\end{equation}
where $\boldsymbol{I}$ is the observed underwater image, $\boldsymbol{J}$ is the real image, $\boldsymbol{A}$ is the attenuation map, and $\boldsymbol{B}$ is the scattering map. While this equation follows the classical underwater physical imaging model framework, our fundamental contribution lies in the sophisticated learning and optimization of the attenuation map $\boldsymbol{A}$ and scattering map $\boldsymbol{B}$.  Our dual modelling mechanism avoids parameters interfering with each other by decoupling the optimization process of attenuation and scattering, and at the same time, introduces a variety of feature-guided strategies to enhance the model expression, RGB image-guided attenuation optimization and depth information-guided scattering optimization, which can accurately simulate optical phenomena in underwater environments, and significantly improve the reconstruction quality and physical accuracy.
\subsubsection{Attenuation Module}
Wavelength-selective attenuation is a key challenge in underwater scene reconstruction.\ Due to the significant difference in the absorption rates of different wavelengths of light in the water body, red light (long wavelength) is rapidly absorbed at the early stage of propagation, resulting in a typical blue-green color tone of distant objects.\ In contrast, blue light (short wavelength) has a stronger penetration ability and can maintain its relative intensity at a longer distance.\ To address this phenomenon, we design an RGB-guided attenuation model that introduces edge feature information of RGB images on top of depth information to enhance the processing capability of complex boundary regions and mitigate the boundary artifacts caused by the discontinuity of the depth map through the edge-aware mechanism.

Our proposed attenuation model enhance depth information utilization rather than simply introducing it, including RGB image features and an edge-aware mechanism, which can be expressed as:
\begin{equation}
\begin{split}
\boldsymbol{A}(\boldsymbol{D}, \boldsymbol{I}_{\mathrm{RGB}}) = &\mathrm{exp} \left( -\sum_{c \in \{\mathrm{r},\mathrm{g},\mathrm{b}\}}  \boldsymbol{w}_c \cdot \beta_c(\boldsymbol{D}) \right. \\
& \left. \cdot (\mathrm{1} - \gamma \cdot E(\boldsymbol{I}_{\mathrm{RGB}}, \boldsymbol{D})) \cdot T(\boldsymbol{I}_{\mathrm{RGB}}) \cdot \boldsymbol{D} \right)
\end{split}
\end{equation}
where $\boldsymbol{w}_{c}$ is the wavelength-dependent weight vector, $\beta_c(\boldsymbol{D})$ is the base attenuation coefficient, $E(\boldsymbol{I}_{\mathrm{RGB}},\boldsymbol{D})$ is the edge perception factor, and $\gamma$ is the edge modulation strength constant. The RGB-guided attenuation model first extracts the color distribution and texture information from the input image through the designed RGB extraction network. This enables the attenuation model to identify color distortion regions in the scene due to wavelength-selective attenuation.\ Combining this RGB information with depth features, the model is able to estimate the actual color applied to each region more accurately.
\subsubsection{Scattering Module}
The scattering intensity increases exponentially with depth as light propagates through a homogeneous body of water, which can be expressed simply as:
\begin{equation}
\boldsymbol{B} = B_{\infty} (\mathrm{1} - \mathrm{e}^{-\beta \cdot d})
\end{equation}
where $\boldsymbol{B}$ denotes the scattering map, $B_{\infty}$ denotes the background color at infinity, $\beta$ denotes the scattering coefficient, and $d$ denotes the depth.

Scattering effects induced by underwater suspended particles are an important cause of degradation of reconstruction quality, and they produce complex angle-dependent scattering of incident light. In addition, unlike homogeneous atmospheric environments, water scattering is highly anisotropic, and scattering from near- and far-field objects behaves differently, making it difficult for a single-scale model to accurately deal with both near-field details and far-field overall characteristics at the same time.

We construct a multi-scale depth-aware scattering model that introduces local structural features and captures depth information at different scales through a feature pyramid network.\ Our physics-guided feature networks are specifically designed for underwater optical physics modeling, extracting features with clear physical meanings including blue-green channel ratios, edge gradients, and color saturation distributions that directly correspond to water body types and optical parameters. These features have demonstrated robust performance through \cref{label:label2}. The model contains three feature extraction branches dealing with original resolution, 1/2 and 1/4 resolution. The multi-scale features are computed as:
\begin{equation}
\begin{aligned}
\boldsymbol{M}(\boldsymbol{D}) = f_{\mathrm{att}}\Big(\
\mathrm{Concat}\big(
&f_{\mathrm{1}}(\boldsymbol{D}), \\
&f_{\mathrm{2}}(\boldsymbol{D}_{\downarrow \mathrm{2}})_{\uparrow \mathrm{2}},
f_{\mathrm{3}}(\boldsymbol{D}_{\downarrow \mathrm{4}})_{\uparrow \mathrm{4}}
\big)\Big)
\end{aligned}
\end{equation}
where $\boldsymbol{D}_{\downarrow n}$ denotes down-sampling $n$ times, $(\cdot)_{\uparrow n}$ denotes upsampling $n$ times, $f_{i}$ denotes the feature extraction function at different scales, and $f_{\mathrm{att}}$ denotes the attention enhancement module. This design is able to focus on both local details and global structure. In order to enhance the feature extraction, we introduce a dual attention mechanism of channel and space in the multi-scale scattering model.\ Channel attention highlights key feature channels through adaptive weighting. In contrast, spatial attention focuses on regions in the image with significant scattering variations, enhancing the model's ability to adapt to different depths and water conditions in different scenes.\ The near-field region relies on high-resolution branches to capture fine edge and texture details. In contrast, the far-field region relies on low-resolution branches to capture the overall scattering trend. The computational formula of our proposed multi-scale scattering model can be expressed as:
\begin{equation}
\begin{split}
\boldsymbol{B}(\boldsymbol{D}) &= B_{\infty} \cdot \left( \mathrm{1} - \mathrm{exp} \left( -\beta_b(\boldsymbol{D}) \cdot C(\boldsymbol{D}) \right. \right. \\
&\left. \left. \cdot \left( \mathrm{1} - \lambda \cdot E_d(\boldsymbol{D}) \right) \cdot \left( \mathrm{1} + \delta \cdot \boldsymbol{M}(\boldsymbol{D}) \cdot \boldsymbol{D} \right) \right) \right)
\end{split}
\end{equation}
where $\beta_{b}(\boldsymbol{D})$ is the base scattering coefficient, $C(\boldsymbol{D})$ is the depth confidence factor, $E_{d}(\boldsymbol{D})$ is the depth edge factor, $\boldsymbol{M}(\boldsymbol{D})$ is the multiscale feature weights, and $\lambda$ and $\delta$ are the modulation factors.\ In addition, we design a depth confidence assessment module to dynamically adjust the scattering calculation according to the depth estimation reliability, and mitigate the boundary artifacts caused by depth discontinuity through the edge-aware processing mechanism, so as to realize the accurate simulation of underwater scattering phenomena.
\subsubsection{Loss Function}
In addition to the basic loss functions, we design attenuation-scattering consistency loss, water body type adaptive loss, edge-aware scattering loss, and multiscale feature loss to ensure that our attenuation and scattering models achieve high-quality reconstruction under a wide range of complex conditions.

In underwater physical imaging models, both attenuation and scattering exhibit specific depth-dependent physical laws, where the scattering effect of the aqueous medium is enhanced with increasing depth, and the attenuation of directly transmitted light intensifies.\ However, previous methods usually optimize scattering and attenuation as independent parameters and lack a physical constraint mechanism. This results in unphysical phenomena of high scattering and low attenuation for close objects, or low scattering and high attenuation for far objects. This inconsistency seriously affects the visual realism, so we devise the attenuation scattering consistency loss:
\begin{equation}
L_{\mathrm{ab}} = \mathbb{E}[\boldsymbol{B}_s \cdot \boldsymbol{D}] - \mathbb{E}[\boldsymbol{A} \cdot \boldsymbol{D}] + \mu \cdot \mathrm{MSE}(\boldsymbol{B}_s + \boldsymbol{T}, \mathrm{1})
\end{equation}
where $\mathbb{E}$ denotes the expectation value, $\boldsymbol{B}_{s}$ denotes the scattering component, $\boldsymbol{D}$ denotes the depth map, $\boldsymbol{A}$ denotes the attenuation component, and $\boldsymbol{T}$ denotes the transmittance map, i.e., the proportion of the original light that successfully passes through the medium and reaches the camera in the underwater scene. This consistency constraint consists of the following three core components, a positive correlation constraint between the scattering term $\boldsymbol{B}_{s}$ and the depth $\boldsymbol{D}$, a negative correlation constraint between the attenuation term $\boldsymbol{A}$ and the depth $\boldsymbol{D}$, and a constraint on the complementarity of the scattering and transmittance $\boldsymbol{T}$. With the above three physical consistency constraints, we ensure that the rendered image is more consistent with the underwater imaging laws at the physical level, enhancing the overall visual realism.

Different water body types have significantly different optical properties, in turbid water bodies, the scattering effect is more significant, while in clear water bodies, the attenuation effect is more prominent.\ In order to adapt to this difference, we designed a water body type adaptive loss function, which estimates the water body type index by analyzing the color distribution characteristics of the image (mainly the ratio of blue and green channels) and then dynamically adjusts the weights of each part of the loss function to achieve adaptive optimization for different water body environments. This adaptive mechanism eliminates the need for manual parameter tuning when applying our method to new datasets. The system automatically identifies water types and adjusts optimization strategies accordingly, ensuring consistent superior performance across diverse underwater environments ranging from clear coral reefs to turbid coastal waters without requiring intervention. The expression can be expressed as:
\begin{equation}
L_{\mathrm{wat}} = \alpha(w) \cdot L_{\mathrm{attenuation}} + \beta(w) \cdot L_{\mathrm{scattering}} + \gamma \cdot L_{\mathrm{ab}}
\end{equation}
where the weighting coefficients $\alpha(w)$ and $\beta(w)$ are dynamically determined by the water body type $w$. We estimate the optical properties of the current water body by mapping the blue-to-green channel ratio of the image to a standardized water body type indicator $w$. In clear water bodies, the model increases the weight of attenuation loss, while in turbid water bodies, it enhances the weight of scattering loss.\ This mechanism allows the training process to adapt to different water body conditions, improving the robustness and accuracy of the model's reconstruction in multiple environments.

In underwater scenes, the edges of the scene usually correspond to depth-continuous regions.\ The scattering characteristics of these regions are significantly different from those of the depth-continuous regions, which makes it difficult for conventional scattering models to accurately deal with the scattering discontinuities at the edges, leading to problems such as blurred edges and distorted structures, which seriously affects the visual realism and geometric accuracy of the rendered image, especially in scenes with complex structures such as coral reefs, and so forth. Especially in scenes with complex structures such as coral reefs.\ To this end, we design a depth-aware edge loss to impose constraints on the imaging characteristics of the scattering and attenuation edge regions to effectively improve the imaging quality and geometric structure preservation ability at the edges. The depth-aware edge loss expression is given by:
\begin{equation}
\begin{split}
L_{\mathrm{edge}} = L_{\mathrm{1}}(\boldsymbol{B}_s \cdot \boldsymbol{W}_{\mathrm{edge}}, &\mathrm{0}) + \lambda \cdot \sum_p \boldsymbol{W}_{\mathrm{smooth}}(p) \\
&\cdot (|\nabla_x \boldsymbol{B}_s(p)| + |\nabla_y \boldsymbol{B}_s(p)|)
\end{split}
\end{equation}
where $\boldsymbol{W}_{\mathrm{edge}}$ is the depth edge weight, which is dynamically assigned according to the magnitude of the local depth gradient, $\boldsymbol{W}_{\mathrm{smooth}}(p) = \mathrm{e}^{-\alpha \cdot (|\nabla_x d(p)| + |\nabla_y d(p)|)}$ is the smoothing constraint weight, which is used to maintain the sharpness of the edges of the structure.\ This loss function imposes stronger constraints in the edge region where the depth varies drastically, prompting the scattering model to learn accurate scattering properties at the edges instead of a simple smoothing process.\ The mechanism pays special attention to the imaging performance at the junction of the foreground object contour and the background water body.\ It can accurately model the scattering transition between the foreground and the background, effectively mitigating the problems of edge blurring and structural distortions, thus significantly improving the edge fidelity in the reconstruction of underwater scenes.

Since it is difficult to capture high-frequency details and low-frequency structural information in an image at a single scale, we propose a multi-scale feature loss function.\ The function captures global structural and local detail information by comparing the model output with the target image at multiple spatial scales. The expression is as follows:
\begin{equation}
L_{\mathrm{ms}} = L_{\mathrm{1}}(\boldsymbol{O}, \boldsymbol{T}) + \frac{\lambda_{\mathrm{ms}}}{|\boldsymbol{S}|} \sum_{s \in \boldsymbol{S}} L_{\mathrm{1}}(D_s(\boldsymbol{O}), D_s(\boldsymbol{T}))
\end{equation}
where $\boldsymbol{O}$ is the rendered image, $\boldsymbol{T}$ is the real image, $\boldsymbol{S}$ is a set of scale factors for downsampling, including the original resolution, $\mathrm{1}/\mathrm{2}$ and $\mathrm{1}/\mathrm{4}$ resolution, $\lambda_{\mathrm{ms}}$ is the weight coefficients, and $D_{s}$ denotes the downsampling processing of the input image and the rendered image at scale $s$.\ $L_{\mathrm{1}}(\boldsymbol{O}, \boldsymbol{T})$ denotes the computation of the $L_{\mathrm{1}}$ loss under the original resolution, and the subsequent terms denote the computation of the loss under the various downsampling scales. By combining the structural losses at different scales, this loss function can maintain consistency at the global layout and local detail levels simultaneously, thus effectively improving the structural fidelity of the rendering results.

Finally, our total loss function expression is:
\begin{equation}
\begin{split}
L_{\mathrm{total}} = w_{\mathrm{1}} L_{\mathrm{basic}} &+ w_{\mathrm{2}} L_{\mathrm{ab}}\\
&+ w_{\mathrm{3}} L_{\mathrm{wat}} 
+ w_{\mathrm{4}} L_{\mathrm{edge}} + w_{\mathrm{5}} L_{\mathrm{ms}}
\end{split}
\end{equation}

\subsubsection{Scene Adaptive Mechanism}
We propose a complete set of scene adaptive mechanisms, which enables DualPhys-GS to dynamically adjust its optimization strategy according to the optical characteristics of different water environments, thus adapting to diverse underwater scenes.

Different water bodies have significantly different optical properties, and it isn't easy to adapt to diverse underwater environments if a fixed single-parameter model is used. Therefore, we developed a classifier to automatically recognize the water body type (clear, medium, turbid) based on the input RGB image features with the following expression:
\begin{equation}
\begin{split}
T(\boldsymbol{I}_{\mathrm{RGB}}) = &\mathrm{Softmax}\\
&\left( \boldsymbol{W}_{\mathrm{2}} \cdot \mathrm{ReLU} \left( \boldsymbol{W}_{\mathrm{1}} \cdot \mathrm{AvgPool}(\boldsymbol{I}_{\mathrm{RGB}}) \right) \right)  
\end{split}
\end{equation}
where $\boldsymbol{I}_{\mathrm{RGB}}$ is the input RGB image, the classifier can evaluate the optical properties of the underwater environment in real-time and realize adaptive processing for different underwater environments.

Different water environments require targeted optimization strategies to achieve optimal reconstruction results. In order to realize the environment-adaptive optimization process, we designed an adaptive optimization controller to dynamically adjust the training strategy according to the water body attributes identified by the water body type classifier.\ The controller mainly consists of the following two functional modules, learning rate adjustment, for the clear water environment, the learning rate is automatically reduced to promote finer parameter convergence. loss function weight adjustment, using the water body type adaptive loss function, the weights of the scattering and attenuation losses are dynamically assigned based on the water body clarity, and the weight of the scattering constraints is increased in the turbid water. in the clear water, the focus is on the attenuation constraints. Through the synergy between the water body type classifiers and the adaptive optimization controllers, we realize the adaptive optimization strategy in diverse underwater environments, significantly improving the model's generalization ability and reconstruction accuracy in different scenarios.

\begin{table*}[!t]
\scriptsize
\centering
\setlength{\tabcolsep}{1.5pt}
\caption{Quantitative comparisons. Qualitative results of the evaluation of the proposed method on the SeaThru-NeRF dataset \cite{levy2023}. "↑" indicates that larger values are better, while '↓' is the opposite (smaller values are better). Values in \textcolor{red}{\textbf{red}} indicate the best results, and values in \textcolor{green}{\textbf{green}} represent the second-best results within the 3DGS category.}
\label{label:label1}
\begin{tabular}{l|l|ccc|ccc|ccc|ccc|ccc}
\toprule
\multirow{2}{*}{Category} & \multirow{2}{*}{Method} & \multicolumn{12}{c|}{SeaThru-NeRF} & \multicolumn{3}{c}{SaltPond} \\
\cmidrule{3-17}
& & \multicolumn{3}{c|}{Curasao} & \multicolumn{3}{c|}{Japanese Gardens} & \multicolumn{3}{c|}{IUI3} & \multicolumn{3}{c|}{Panama} & \multicolumn{3}{c}{SaltPond} \\
& & PSNR↑ & SSIM↑ & LPIPS↓ & PSNR↑ & SSIM↑ & LPIPS↓ & PSNR↑ & SSIM↑ & LPIPS↓ & PSNR↑ & SSIM↑ & LPIPS↓ & PSNR↑ & SSIM↑ & LPIPS↓ \\
\midrule
\multirow{1}{*}{NeRF} & Seathru-NeRF \cite{levy2023} & 30.08 & 0.87 & 0.19 & 21.74 & 0.77 & 0.29 & 26.01 & 0.79 & 0.32 & 27.69 & 0.83 & 0.28 & 11.93 & 0.51 & 0.58 \\
\midrule
\multirow{1}{*}{NeRF+3DGS} & WaterSplatting \cite{li2024watersplatting} & 32.67 & 0.96 & 0.11 & 25.20 & 0.90 & 0.11 & 29.39 & 0.91 & 0.18 & 30.49 & 0.95 & 0.08 & 26.83 & 0.66 & 0.38 \\
\midrule
\multirow{4}{*}{3DGS} 
& 3DGS \cite{kerbl2023} & 28.01 & 0.88 & 0.21 & 21.47 & 0.85 & 0.22 & 21.11 & 0.81 & 0.29 & 29.64 & 0.90 & 0.17 & 27.10 & \textcolor{green}{\textbf{0.75}} & \textcolor{green}{\textbf{0.29}} \\
& Seasplat \cite{yang2024} & 30.30 & 0.90 & \textcolor{green}{\textbf{0.19}} & 22.70 & \textcolor{red}{\textbf{0.87}} & \textcolor{red}{\textbf{0.18}} & 26.67 & \textcolor{green}{\textbf{0.87}} & \textcolor{green}{\textbf{0.21}} & 28.76 & 0.90 & 0.15 & \textcolor{green}{\textbf{27.47}} & \textcolor{green}{\textbf{0.75}} & \textcolor{red}{\textbf{0.25}} \\
& UW-GS\cite{wang2025uwgs} & \textcolor{red}{\textbf{31.77}} & \textcolor{red}{\textbf{0.94}} & \textcolor{red}{\textbf{0.14}} & \textcolor{red}{\textbf{23.05}} & \textcolor{green}{\textbf{0.86}} & \textcolor{green}{\textbf{0.19}} & \textcolor{red}{\textbf{28.65}} & \textcolor{red}{\textbf{0.93}} & \textcolor{red}{\textbf{0.13}} & \textcolor{red}{\textbf{31.79}} & \textcolor{red}{\textbf{0.94}} & \textcolor{red}{\textbf{0.12}} & - & - & - \\
& Ours & \textcolor{green}{\textbf{30.48}} & \textcolor{green}{\textbf{0.91}} & \textcolor{green}{\textbf{0.19}} & \textcolor{green}{\textbf{22.77}} & \textcolor{red}{\textbf{0.87}} & \textcolor{red}{\textbf{0.18}} & \textcolor{green}{\textbf{27.86}} & \textcolor{green}{\textbf{0.87}} & 0.23 & \textcolor{green}{\textbf{29.9}} & \textcolor{green}{\textbf{0.91}} & \textcolor{green}{\textbf{0.14}} & \textcolor{red}{\textbf{28.03}} & \textcolor{red}{\textbf{0.77}} & \textcolor{red}{\textbf{0.25}} \\
\bottomrule
\end{tabular}

\end{table*}

\begin{figure*}[!t]%% placement specifier
%% Use \includegraphics command to insert graphic files. Place graphics files in 
%% working directory.
\centering%% For centre alignment of image.
\includegraphics[width=\textwidth]{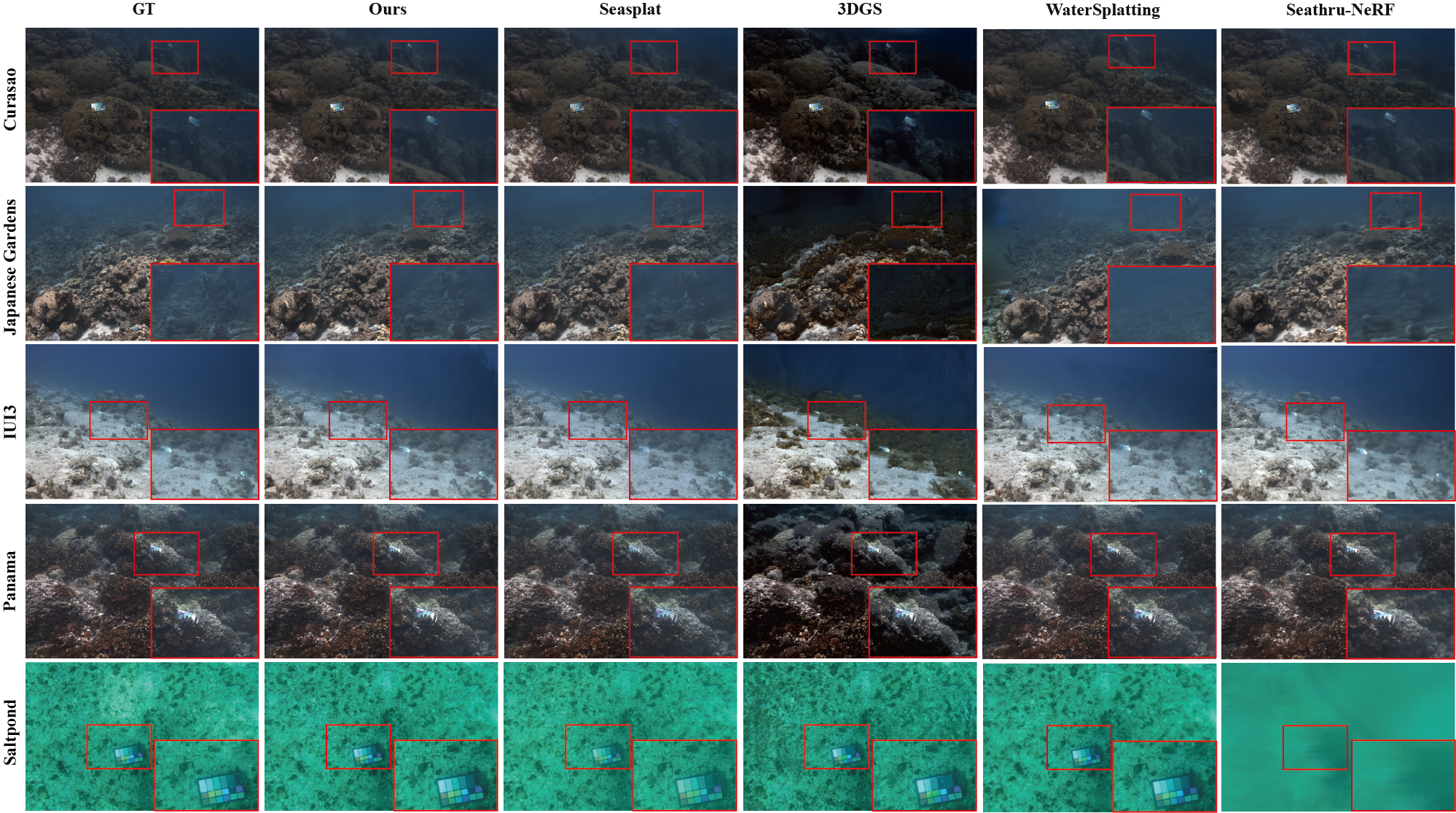}
%% Use \caption command for figure caption and label.
% \captionsetup{justification=raggedright, singlelinecheck=false}
\caption{Comparison of novel synthesis methods based on underwater media. Dualphys-GS exhibits high-quality rendering results.}\label{fig:fig2}
%% <url id="d0jkdj4chmtjjc0390mg" type="url" status="failed" title="" wc="0">https://en.wikibooks.org/wiki/LaTeX/Importing_Graphics#Importing_external_graphics</url> 
\end{figure*}

\section{Experiments}
\subsection{Datasets}
To validate the effectiveness of the proposed method, we use the multi-view underwater scene dataset published by SeaThru-NeRF \cite{levy2023} and the SaltPond dataset for experimental evaluation. The SeaThru-NeRF dataset \cite{levy2023} covers typical underwater scenes from multiple sea areas, including four sub-scenes: Japanese Gardens, IUI3, Curasao, and Panama.\ These scenes are rich in depth variations and optical diversities and can comprehensively reflect the imaging characteristics in different water body environments.

\begin{figure*}[!t]%% placement specifier
%% Use \includegraphics command to insert graphic files. Place graphics files in 
%% working directory.
\centering%% For centre alignment of image.
\includegraphics[width=\textwidth]{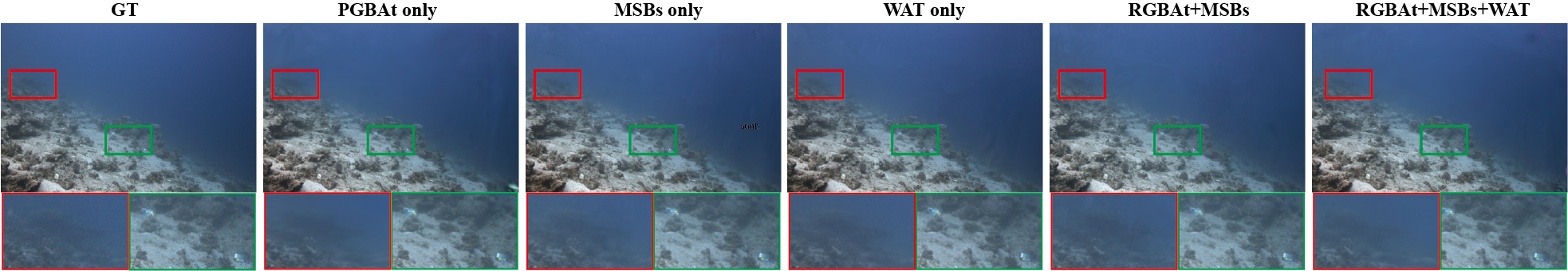}
%% Use \caption command for figure caption and label.
% \captionsetup{justification=raggedright, singlelinecheck=false}
\caption{Model Ablations in \textbf{IUI3-RedSea} from Seathru-NeRF dataset \cite{levy2023}. We have shown partial details of rendered images with progressive module integration.}\label{fig:fig3}
%% <url id="d0jkdj4chmtjjc0390mg" type="url" status="failed" title="" wc="0">https://en.wikibooks.org/wiki/LaTeX/Importing_Graphics#Importing_external_graphics</url> 
\end{figure*}

\begin{figure*}[!t]%% placement specifier
%% Use \includegraphics command to insert graphic files. Place graphics files in 
%% working directory.
\centering%% For centre alignment of image.
\includegraphics[width=\textwidth]{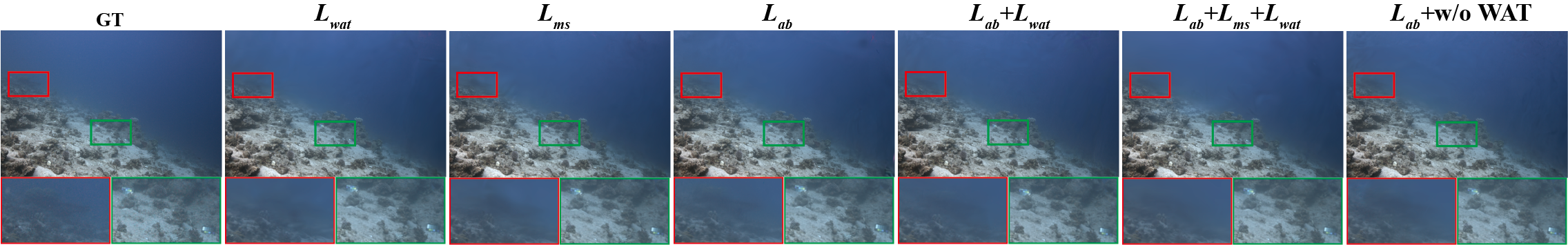}
%% Use \caption command for figure caption and label.
% \captionsetup{justification=raggedright, singlelinecheck=false}
\caption{Loss Ablations in \textbf{IUI3-RedSea} from Seathru-NeRF dataset \cite{levy2023}. We have shown partial details of rendered images with progressive loss function integration.}\label{fig:fig4}
%% <url id="d0jkdj4chmtjjc0390mg" type="url" status="failed" title="" wc="0">https://en.wikibooks.org/wiki/LaTeX/Importing_Graphics#Importing_external_graphics</url> 
\end{figure*}

\subsection{Evaluation Metrics}
We chose three widely used image evaluation metrics to validate the new perspective synthesis effect.\ We assessed the visual fidelity by comparing the final obtained rendered image with the real image through the peak signal-to-noise ratio (PSNR), the structural similarity index (SSIM) \cite{wang2004image}, and the perceptual image block similarity (LPIPS) \cite{zhang2018unreasonable}.

\subsection{Implementation Details}
Before training begins, we use COLMAP \cite{schonberger2016structure} to get an initialized point cloud and estimate the camera position. The number of training iterations for all scenes is $\mathrm{30{,}000}$, and underwater color rendering is turned on at $\mathrm{10{,}000}$ iterations. In the initialization phase, we designed a scene adaptation mechanism for different underwater environments to automatically recognize five main water types (Curasao, JapaneseGradens-RedSea, IUI3-RedSea, Panama, and SaltPond).\ During the training process, we use the Adam optimizer to optimize the Gaussian representation parameters, scattering model and attenuation model. The learning rate thresholds for the attenuation and scattering models are set to $\mathrm{5e{-}4}$ and $\mathrm{1e{-}4}$, respectively, with optimization strategies adjusted by our scene-adaptive mechanism for different water body environments. For water bodies classified as "clear" (e.g., Curasao), the learning rates of the attenuation and scattering models are dynamically reduced from $\mathrm{1e{-}4}$ to $\mathrm{5e{-}5}$ for fine optimization, while the water body type adaptive loss $L_{\mathrm{wat}}$ increases the weight of the attenuation loss $\alpha(w)$ to $\mathrm{1.2}$ and decreases the weight of the scattering loss $\beta(w)$ to $\mathrm{0.8}$. Additionally, in "turbid" water bodies (e.g., IUI3), the learning rate is maintained at $\mathrm{1e{-}4}$ to accelerate convergence, while the weight of scattering loss $\beta(w)$ is increased to $\mathrm{1.2}$ and the weight of attenuation loss $\alpha(w)$ is reduced to $\mathrm{0.8}$ to strengthen constraints on scattering effects. This automated parameter adjustment strategy is the key to ensuring the effectiveness of our method across diverse datasets. All experiments in this paper are done on a single workstation configured with an Intel Core i9-13900K processor, 64GB DDR5 memory, and an NVIDIA GeForce RTX 4090 graphics card (24GB video memory).

% \begin{table}[!t]
% \small
% \centering
% % \setlength{\tabcolsep}{5pt}
% \begin{tabular}{l|c|c|c}
% \toprule
% Method & Training & Rendering & FPS \\
% & Time (h) & Time (ms) & \\
% \midrule
% Watersplatting \cite{li2024watersplatting} & 8.3 & 125.6 & 8.0 \\
% 3DGS \cite{kerbl2023} & 0.8 & \textcolor{green}{\textbf{15.2}} & \textcolor{green}{\textbf{65.8}} \\
% SeaSplat\cite{yang2024} & 1.2 & 18.7 & 53.5 \\
% Ours & \textcolor{red}{\textbf{0.9}} & \textcolor{red}{\textbf{12.8}} & \textcolor{red}{\textbf{78.1}} \\
% \bottomrule
% \end{tabular}
% \caption{Comprehensive time and performance comparison.}
% \label{tab:time_comparison}
% \end{table}

\subsection{Results and Discussion}
\subsubsection{Quantitative results}
Table \ref{label:label1} presents a comprehensive evaluation of different methods on the SeaThru-NeRF dataset~\cite{levy2023} and the SaltPond dataset. In terms of quantitative metrics, WaterSplatting~\cite{li2024watersplatting} demonstrates a significant advantage in PSNR on the SeaThru-NeRF dataset. This primarily stems from its regularization loss function, which more directly fits the visual characteristics of target images. Notably, while our method exhibits a performance gap in pure numerical metrics, we prioritize fidelity in physical process modeling over solely pursuing visual feature fitting. In addition, WaterSplatting  \cite{li2024watersplatting} does not use the 3DGS framework.

Compared to UW-GS~\cite{wang2025uwgs}, which achieves superior PSNR performance on most SeaThru-NeRF scenes, it should be noted that UW-GS leverages the pre-trained depth estimation model DepthAnything~\cite{yang2024depthanything} to enhance depth accuracy. In contrast, our approach relies solely on the depth information rendered by 3DGS without external depth priors, demonstrating the effectiveness of our physics-guided dual modeling strategy under more constrained conditions.

This design distinction manifests in the SaltPond dataset results. DualPhys-GS achieves the highest PSNR value of 28.03 on this dataset, surpassing all comparative methods including WaterSplatting \cite{li2024watersplatting}. This indicates that physics-constrained modeling methods exhibit greater generalization potential when confronted with diverse aquatic environments.

In the Structural Similarity (SSIM)~\cite{wang2004image} assessment, despite the PSNR gap, DualPhys-GS consistently achieves SSIM scores comparable to the best-performing methods across multiple scenes. This suggests an advantage in preserving structural image integrity for our approach, likely attributed to its physics-based modeling properties.

The perceptual quality metric (LPIPS)~\cite{zhang2018unreasonable} further supports the effectiveness of our method. In scenes such as Curasao and Japanese Gardens, DualPhys-GS yields perceptual quality on par with top methods. This implies that despite pixel-level accuracy differences, the reconstruction results from our method exhibit comparable perceptual acceptability from a human visual perspective.

Regarding computational efficiency, DualPhys-GS demonstrates a substantial advantage. Its rendering time of 0.016s represents an approximately 5-fold improvement over WaterSplatting's 0.084s.
\subsubsection{Qualitative results}
Regarding visual reconstruction quality, the comparison of the rendering results in \Cref{fig:fig2} reveals the differences in the characteristics of the different methods. Our proposed DualPhys-GS method performs well in terms of reconstruction quality due to its innovative dual-path optimization mechanism.

Traditional 3DGS \cite{kerbl2023} methods often suffer from unstable Gaussian distributions in scenes with complex geometries, such as the coral reefs in Panama.\ In contrast, our method can accurately recover the color information of distant objects through the RGB-guided attenuation model, especially the long-wavelength red channel, which successfully solves the problem of distant objects' unnatural bluish-green hue.

At the same time, the multi-scale scattering model can accurately capture the scattering effects at different scales, which enables the model to generate smooth and continuous depth estimation in the underwater environment while preserving the detailed features.\ Furthermore, the water type adaptation mechanism enables the model to automatically adjust the optimization strategy according to different water environments, thus maintaining excellent performance under various water conditions and demonstrating high adaptability to complex underwater environments.
\begin{table}[!t]
\small
\centering
\caption{Comparison of rendering times for different methods on the Seathru-NeRF dataset.}
\label{tab:time_comparison}
\begin{tabular}{l|c}
\toprule
Method  & Rendering Time  \\

\midrule
Watersplatting \cite{li2024watersplatting}  & 0.084 s  \\
3DGS \cite{kerbl2023} & 0.006 s \\
SeaSplat\cite{yang2024} & 0.012 s  \\
Ours & 0.016 s \\
\bottomrule
\end{tabular}

\end{table}
\begin{table}[!t]%% placement specifier
\small
%% Use tabular environment to tag the tabular data.
%% https://en.wikibooks.org/wiki/LaTeX/Tables#The_tabular_environment
\centering%% For centre alignment of tabular.
\caption{Ablations. We measure the average values of all metrics across the Seathru-NeRF dataset, where RGBAt is RGB-guided attenuation optimization model, MSBs is multi-scale depth-aware scattering model and WAT is water body scene adaptive module based on RGB guidance and loss function.}\label{label:label2}
\begin{tabular}{l|ccc}
        \toprule
        Metrics & PSNR$\uparrow$ & SSIM$\uparrow$ & LPIPS$\downarrow$ \\
        \midrule
        Total & 27.63 & 0.89 & 0.19 \\
        \midrule
        Model & & & \\
        RGBAt only & 26.86 & 0.88 & 0.19 \\
        MSBs only & 27.07 & 0.89 & 0.19 \\
        WAT only & 27.20 & 0.89 & 0.18 \\
        RGBAt+MSBs & 27.27 & 0.89 & 0.18 \\
        RGBAt+MSBs+WAT & 27.29 & 0.89 & 0.18 \\
        \midrule
        Loss & & & \\
         $L_{\mathrm{wat}}$ only & 27.08 & 0.88 & 0.18 \\
         $L_{\mathrm{ms}}$ only & 27.14 & 0.89 & 0.19 \\
         $L_{\mathrm{ab}}$ only & 27.39 & 0.89 & 0.19 \\
         $L_{\mathrm{ab}}$+$L_{\mathrm{wat}}$ & 27.40 & 0.89 & 0.18 \\
         $L_{\mathrm{ab}}$+$L_{\mathrm{wat}}$+$L_{\mathrm{ms}}$ & 27.57 & 0.89 & 0.19 \\
         $L_{\mathrm{ab}}$+w/o WAT & 27.47 & 0.89 & 0.18 \\
        \bottomrule
\end{tabular}
%% Use \caption command for table caption and label.
% \captionsetup{justification=raggedright, singlelinecheck=false}

\end{table}

\subsection{Ablation Study}
Through \Cref{label:label2}, we systematically evaluate the effectiveness of key components in DualPhys-GS, including feature-guided attenuation-scattering modeling, aquatic scene adaptation module, and analyzing the effects of different loss functions.\ Our experiments are conducted on the SeaThru-NeRF dataset, with comparative analysis of average PSNR, SSIM, and LPIPS values across four scenes to elucidate the impacts of individual modules and loss terms on model performance.
\subsubsection{Model Ablation}
Quantitative results from \Cref{label:label2} and visualizations in \Cref{fig:fig3} demonstrate that the complete DualPhys-GS model achieves optimal reconstruction quality. In contrast, using only the RGB feature-based attenuation model or the multi-scale depth-aware scattering model individually results in compromised performance.\ This validates that our dual-branch modeling mechanism effectively integrates RGB features with depth information to address underwater challenges including long-range color distortion and wavelength-selective attenuation, while the multi-scale scattering modeling employs feature pyramid architecture to capture scattering effects across scales, thereby enhancing detail preservation.

Furthermore, the water-body-type adaptation module contributes significantly to performance improvement.\ When activated independently, this module dynamically adjusts optimization strategies according to water characteristics, outperforming single-module configurations. The combination of RGB attenuation and multi-scale scattering modules achieves incremental quality gains, with the full model exhibiting synergistic effects that maximize complementary advantages among components, ultimately attaining best results.
\subsubsection{Loss Ablation}
Our ablation studies further validate the importance of individual loss functions.\ The visual details in \Cref{fig:fig4} demonstrate concrete improvements, while \Cref{label:label2} quantitatively reveals that the attenuation-scattering consistency loss plays a pivotal role in performance enhancement.\ When exclusively employing this loss, the model achieves superior reconstruction quality compared to using only scene-adaptive loss or multi-scale feature loss.\ This confirms that physical consistency constraints (e.g., positive scattering-depth correlation, negative attenuation-depth relationship, and scattering-transmittance complementarity) are vital for accurate underwater scene reconstruction.

The synergistic combination of attenuation-scattering consistency loss with scene-adaptive loss enables the model to maintain physical plausibility while adapting to diverse aquatic environments. Subsequent integration of multi-scale feature loss strengthens the model's capacity to capture both global structures and local details, yielding incremental quality gains. Notably, even with all three loss terms integrated, the model still exhibits degraded reconstruction when lacking the aquatic-type adaptation mechanism.
\section{Limitations}
Before deploying DualPhys-GS in real-world underwater application scenarios, the method must be adapted and optimized for efficient operation in real-time environments, especially for computational optimization in dynamically constructing 3D Gaussian representations.\ This has potential applications for autonomous underwater navigation and adaptive sampling in special areas such as coral reefs.\ Although our framework accurately models the attenuation and scattering effects of the water column through a dual-path optimization mechanism, there are still some limitations in terms of the completeness of the optical modeling: the focal dispersion phenomenon resulting from refraction at the water surface, as well as the impact of underwater equipment (e.g., divers or camera devices) on the scene illumination, which is particularly noticeable in some datasets (e.g., SaltPond), are not adequately taken into account. In addition, real underwater environments are full of dynamic elements (e.g., seaweed swaying with the current or fish swimming). DualPhys-GS is currently optimized for static underwater scenes, so the ability to model dynamic underwater scenes or objects needs to be further enhanced.
\section{Conclusions}
In this paper, we propose a dual-path optimization framework DualPhys-GS based on 3DGS \cite{kerbl2023} to address the color distortion and geometric artifacts in 3D reconstruction of underwater scenes, and we achieve an accurate simulation of underwater optical propagation through a feature-guided attenuation-scattering dual modeling mechanism. The RGB-guided attenuation optimization model combines RGB features and depth information to deal with scene boundaries and structural details accurately. In contrast, the multi-scale depth-aware scattering model captures the optical effects at different scales through a feature pyramid network and an attention mechanism. A series of loss functions (e.g., edge-aware scattering loss, multi-scale feature loss, attenuation scattering consistency loss, etc.) are designed to ensure that the model outputs are consistent with the physical laws of underwater optics. In addition, the scene adaptation mechanism can automatically recognize the water type according to the image features and adjust the parameters accordingly so as to adapt to different underwater environments, from clear to turbid. Experimental results show that DualPhys-GS significantly improves the reconstruction's geometric accuracy and texture fidelity in complex underwater scenes, especially in areas with dense suspended objects and long-distance seabed terrain.
% \section{Acknowledgments}
% Removed for anonimity

\end{document}